# Efficient Electrical Spin Injection from a Magnetic Metal / Tunnel Barrier Contact into a Semiconductor

A. T. Hanbicki,[a] B. T. Jonker,[a] G. Itskos,[b] G. Kioseoglou,[b] and A. Petrou [b]

[a] Materials Physics Branch, Naval Research Laboratory, Washington, DC 20375

[b] Department of Physics, State University of New York, Buffalo, New York 14260

## ABSTRACT

We report electrical spin injection from a ferromagnetic metal contact into a semiconductor light emitting diode structure with an injection efficiency of 30% which persists to room temperature. The Schottky barrier formed at the Fe/AlGaAs interface provides a natural tunnel barrier for injection of spin polarized electrons under reverse bias. These carriers radiatively recombine, emitting circularly polarized light, and the quantum selection rules relating the optical and carrier spin polarizations provide a quantitative, model-independent measure of injection efficiency. This demonstrates that spin injecting contacts can be formed using a widely employed contact methodology, providing a ready pathway for the integration of spin transport into semiconductor processing technology.

The intrinsic spin exhibited by electrons and holes offers new functionality and performance for semiconductor devices,[1,2] as well as an avenue to circumvent the dielectric breakdown and capacitive limits which are major near-term concerns in existing electronics. Efficient electrical injection of spin-polarized carriers from a contact into a semiconductor is one of the essential requirements to utilize carrier spin as a new operational paradigm for future electronic devices.[3] Previous efforts to produce reasonable spin injection efficiencies have been restricted to low temperature [4,5,6] or required the use of a scanning tunneling microscope.[7] We report here that a ferromagnetic metal employed as a conventional Schottky barrier contact provides a practical and robust means of injecting spin polarized carriers into a semiconductor device heterostructure, with spin injection efficiencies of 30% extending to room temperature. This demonstrates that spin injecting contacts can be formed using a very familiar and widely employed contact methodology, and provides a ready pathway for the integration of spin transport into semiconductor processing technology.

Ferromagnetic metals are attractive as spin injecting contacts because they offer high Curie temperatures and enjoy a very high level of material development due to their widespread use in magnetic recording technology. Recent calculations have shown, however, that fundamental considerations preclude their use as efficient spin injectors in the diffusive transport regime (in which all current electronic devices operate), due to the large mismatch in conductivities between the metal and semiconductor.[8] A number of groups have attempted to inject spin polarized carriers from a ferromagnetic metal into a semiconductor, and reported *measured* effects on the order of 0.1-1%, with an estimate of injection efficiency extracted from a particular model.[9,10,11,12] These experiments typically measured a change in resistance or potential, which some argue may be complicated by possible contributions from anisotropic magnetoresistance or a local Hall effect. [13, 14, 15]

A model independent method of determining spin injection efficiency utilizes a spin-polarized light emitting diode (spin-LED), in which spin polarized carriers injected from the contact radiatively recombine in the semiconductor. If these carriers retain their spin polarization, the emitted light is circularly polarized. The quantum selection rules which describe the radiative recombination process provide a direct and fundamental link between the circular polarization of the light emitted along the surface normal, $P_{circ}$, and the spin polarization of the electrically injected carriers, $P_{spin}$.[5, 16] Hence the spin-LED provides a *quantitative* and *model independent* measure of electrical spin injection from any given contact.

This approach has been employed for measuring spin injection efficiency from magnetic semiconductor contacts at low temperature.[4,5,6] The light emitted from these spin-LEDs exhibited large differences in intensity (up to 12-fold) when analyzed as left (σ+) or right (σ−) circularly polarized, corresponding to $P_{circ} = 85\%$.[17] Zhu *et al.* have also utilized this approach to examine spin injection from Fe films grown epitaxially on a GaAs/InGaAs quantum well based structure.[18] Although they observe no large differences in electroluminescence (EL) intensity when analyzed as σ+ and σ− , by examining the wings of the Gaussian-like intensity distribution to distinguish the heavy hole exciton contribution, they concluded that a 2% spin injection efficiency had been realized which was independent of temperature.

We have fabricated spin-LEDs designed to provide efficient *tunnel barrier* injection of spin-polarized electrons from epitaxial Fe film contacts into an AlGaAs/GaAs quantum well (QW) LED structure. The Fe forms a Schottky barrier contact to the AlGaAs, which is tailored to provide a tunnel barrier for electrical injection of spin polarized electrons under reverse bias (Fig. 1a). A tunnel barrier supports a difference in electrochemical potential between spin-up and spin-down carrier bands,[19] and hence circumvents the constraint of mismatched conductivity between the contact and semiconductor described in reference 8. We observe a large difference in the intensity of the EL spectrum when analyzed for $\sigma+$ and $\sigma-$ polarization, corresponding to a spin injection efficiency of 30% based on the quantum selection rule analysis, which persists to near room temperature.

The LED heterostructures were grown by molecular beam epitaxy using interconnected growth chambers. The 125 Å thick Fe(001) film was grown with the substrate at 10-15°C to minimize potential intermixing at the $Al_{0.08}Ga_{0.92}As$ interface. The doping profile of the top 800 Å thick n-type $Al_{0.08}Ga_{0.92}As$ layer was chosen so that the resulting Schottky contact had a narrow depletion width, thus forming a triangular shaped tunnel barrier, as shown in Figure 1(a). This enables spin polarized electrons to tunnel from the Fe into the semiconductor under applied bias (reverse biased Schottky barrier) rather than diffusively transport, as would be the case with an ohmic contact. The width of the GaAs quantum well was chosen to be 100 Å to insure separation of the light and heavy hole levels and corresponding excitonic spectral features, an important consideration for quantitative interpretation of the data.[5] The samples were processed to form surface emitting LEDs using conventional photolithography and chemical etching techniques. The light emitted along the surface normal (Faraday geometry) is analyzed for left ($\sigma+$) and right ($\sigma-$) circular polarization and spectroscopically resolved using a quarter wave plate and linear polarizer followed by a spectrometer. Further details may be found elsewhere.[5]

Electroluminescence spectra obtained from the LED structures are shown in Figure 1(b). The EL spectrum is dominated by a feature at 1.53 eV with a full width at half maximum of 17 meV, due to the heavy hole exciton in the GaAs QW. With no applied magnetic field, the Fe magnetization (easy axis) and corresponding electron spin orientation are entirely in the plane of the thin film. Therefore, the average electron spin along the z-axis is zero, and the σ+ and σ- components are nearly coincident, as expected. The small difference observed corresponds to less than 2% polarization, and is treated as a background in subsequent data analysis. A magnetic field is applied along the surface normal to align the electron spins in the Fe along the z-axis. If these carriers reach the QW with a net spin polarization and radiatively recombine, circularly polarized light is emitted along the surface normal, as described earlier. A simple inspection of the raw data reveals that carrier spin injection indeed occurs: as the magnetic field increases, the component of Fe magnetization and electron spin polarization along the z-axis continuously increase, and the corresponding spectra exhibit a substantial difference in intensity of the σ+ and σ- components which rapidly increases with field.

The circular polarization is defined as $P_{circ} = (I^+ - I^-) / (I^+ + I^-)$, where $I^+$ and $I^-$ are the EL component peak intensities when analyzed as σ+ and σ-, respectively. These data are summarized in Figure 2 as a function of magnetic field. $P_{circ}$ rapidly increases in magnitude until the Fe moment is saturated out-of-plane. This occurs at a field value determined by the magnetic properties of Fe, $4\pi M = 2.2$ T (indicated by the fiducial marks on the top axis in the figure), where $M$ is the Fe magnetization. The polarization continues to increase at higher fields, but at a much slower rate of 1.25% / T, which is attributed to a combination of Zeeman splitting in the GaAs and an instrumental background – this uniform linear background has been subtracted from the data of Figure 2 and all subsequent data presented here. Note that $P_{circ}$ changes sign as the field direction is reversed – the field dependence of $P_{circ}$ mirrors that of the out-of-plane magnetization of the Fe film, obtained by independent superconducting quantum interference device magnetometry measurements and shown as a dot-dashed line. The sign of the polarization[20] demonstrates *minority* spin injection from the Fe Schottky tunnel contact, *i.e.* that the polarization of the injected current corresponds to minority spin in the Fe, in contrast to the model proposed by Stearns.[21] However, recent work has shown that a number of factors are likely to contribute to the spin polarization of the tunneling current, including the interface electronic structure and barrier thickness.[22]

A number of control experiments were performed to rule out spurious effects. LED structures fabricated with the Fe contact removed showed little circular polarization and very weak field dependence. Possible contributions to the measured $P_{circ}$ arising from Faraday rotation as the light emitted from the QW passes through the Fe film were determined both analytically and directly measured, and found to be less than 1%. This contribution was calculated to be 0.9% using well established models at the appropriate wavelength for the thickness of the Fe film.[23] This contribution was also directly measured for the samples included here by measuring the circular polarization of the GaAs QW photoluminescence excited by linearly polarized light and emitted along the surface normal (through the Fe film). Since linearly polarized optical excitation produces zero net carrier spin polarization in the GaAs QW, the emitted light is unpolarized, and any measured circular polarization is derived from Faraday rotation produced by the Fe film. These results are shown as open triangles in Figure 2, and are ~ 1%. Note that the effect measured due to electron spin injection is over an order of magnitude larger.

The quantum selection rules provide a quantitative and model-independent link between the measured $P_{circ}$ and the spin polarization of the carriers which radiatively recombine in the QW, $P_{spin}$, as described above. For the case here in which only the heavy hole levels participate, $P_{circ} = P_{spin}$.[5] The spin injection efficiency, $\eta$, from the Fe contact through the Schottky tunnel barrier and into the QW can be readily calculated as $\eta = P_{spin}/P_{Fe} = P_{circ}/P_{Fe}$, where $P_{Fe}$ is the spin polarization of the Fe contact near the Fermi level. Using $P_{circ} = 13\%$ from Figure 2, and $P_{Fe} = 0.44$,[24] we obtain $\eta = 30\%$ at 4.5 K. This is comparable to the injection efficiencies reported using magnetic semiconductor contacts.[4,5]

$P_{circ}$ decreases with temperature, as shown in Figure 3. EL spectra obtained at 90 K and 240 K at a field of 3T (Fe magnetization saturated normal to the surface) are shown as insets, and exhibit linewidths of 20 and 25 meV, respectively. The circular polarization is 8.5% at 90 K, and 4% at 240 K. To obtain corresponding values for the electrical spin injection efficiency, one must recognize that electron spin relaxation in the GaAs QW itself occurs more rapidly with increasing temperature, suppressing the measured circular polarization. This effect is completely independent of the spin injection process, and must be corrected for. Adachi et al[25] report experimental values for the electron spin relaxation time, $\tau_s$, for the full temperature range considered here. We use these values to provide a first order correction for the spin relaxation occuring in the QW by weighting our measured circular polarization at 90 and 240 K by the ratio of the corresponding value of $\tau_s$ to that at low temperature. This yields values for the electrical spin injection efficiency of 32% and 30% at 90 and 240 K, respectively, as shown by the square symbols in Figure 3. Thus the spin injection efficiency exhibits little temperature dependence, and is approximately 30% as one approaches room temperature.

In summary, the ferromagnetic Schottky contact approach described here provides a very simple tunnel barrier which enables a practical means of injecting spin polarized electrons into a semiconductor at room temperature. This approach does not require a discrete insulating layer (e.g. $Al_2O_3$), and thereby avoids the accompanying complications of pinholes and non-uniform barrier thickness.

This work was supported by the DARPA SpinS program and the Office of Naval Research. The authors thank J. J. Krebs for calculation of the polarization due to Faraday rotation in the Fe film, and M. Furis for assistance with the optical measurements.

# FIGURE CAPTIONS

**Figure 1:** (a) Flat band energy diagram of the Schottky tunnel barrier and LED structure. (b) EL spectra at selected magnetic fields and T = 4.5 K, analyzed for σ+ and σ− circular polarization. The large difference in intensity between these components indicates successful spin injection. Typical operating parameters were 10 mA and 1.8 V at 4.5 K.

**Figure 2:** The field dependence of $P_{circ}$ (solid circles) and the out-of-plane component of the Fe film magnetization (dot-dashed line, normalized to the maximum value of $P_{circ}$). The sign of $P_{circ}$ changes as the Fe magnetization is reversed, and reflects minority electron spin injection from the Fe contact. The triangles show the measured contribution to $P_{circ}$ due to Faraday rotation in the Fe film (1% ± 1%). All data obtained at T = 4.5 K.

**Figure 3:** Temperature dependence of $P_{circ}$ and the corresponding spin injection efficiency (circles). The square symbols show the spin injection efficiency after correction for the unrelated process of spin relaxation in the GaAs quantum well. The insets show the EL spectra obtained at the temperatures indicated (H = 3 T), analyzed for σ+ and σ− polarization. A significant difference in intensity is observed even near room temperature.

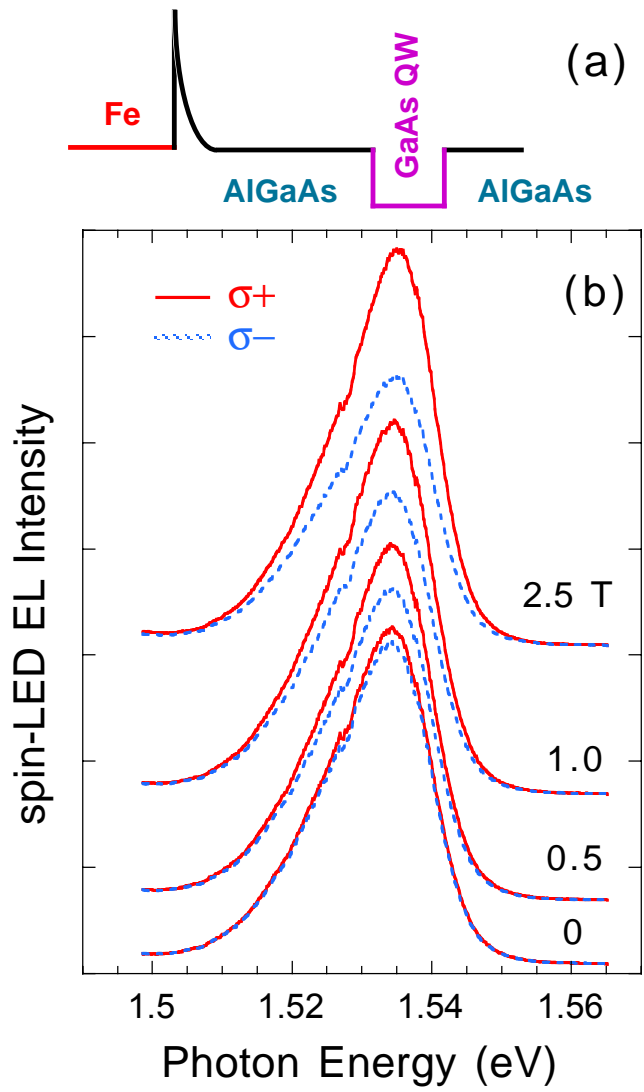

Figure 1 -- Hanbicki, et al.

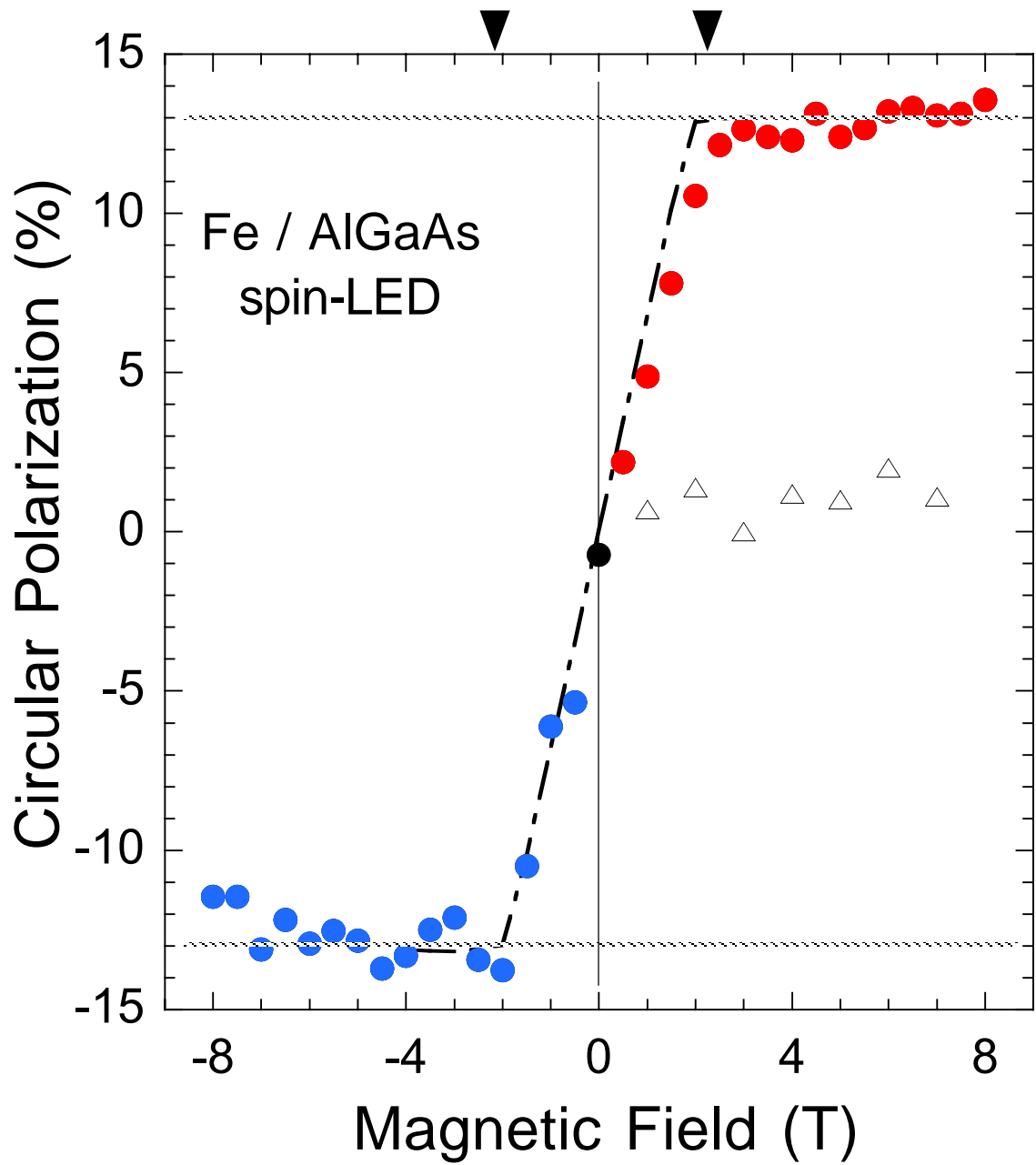

Figure 2 Hanbicki et al

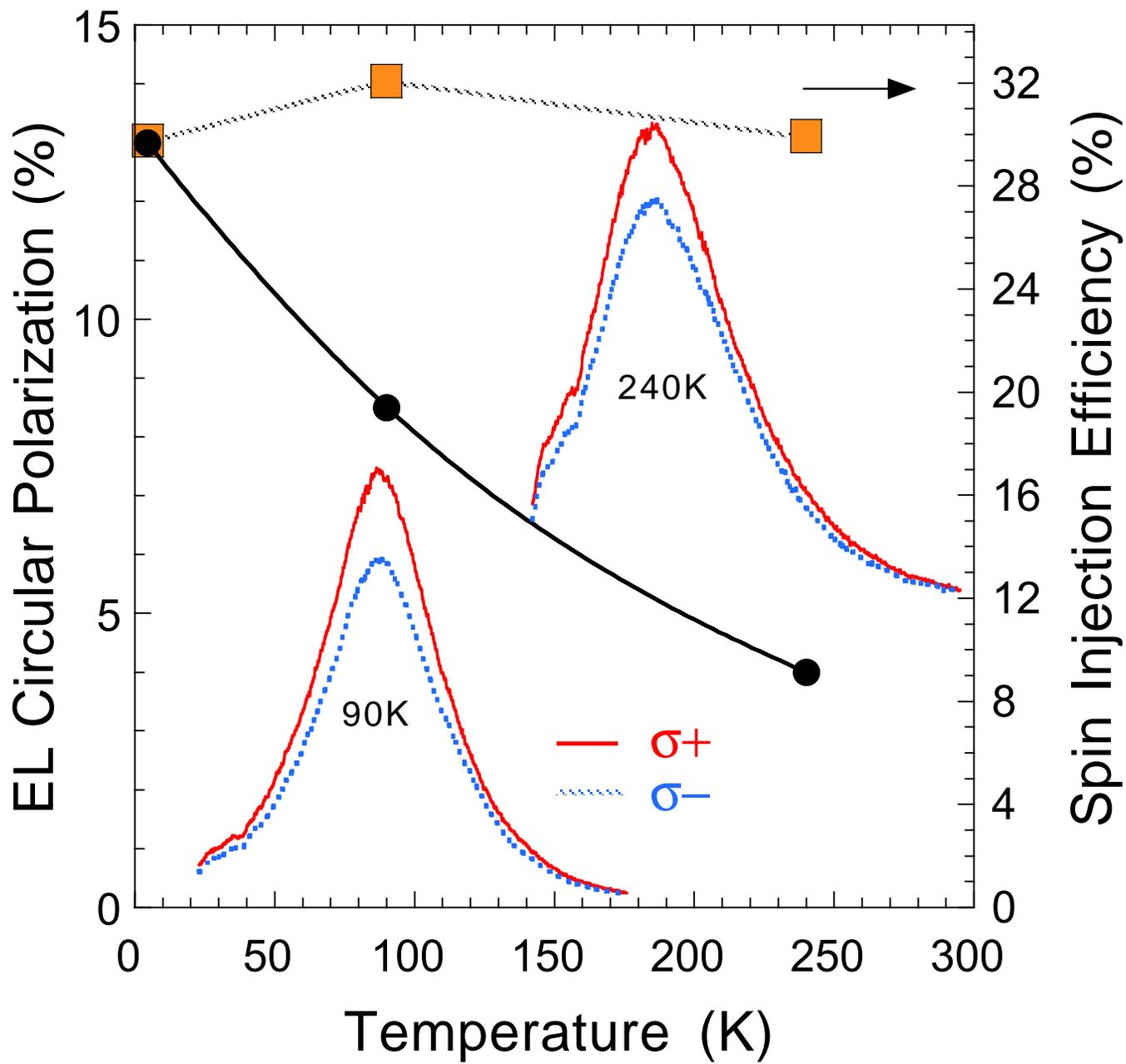

Figure 3 Hanbicki et al